\documentclass[12pt,a4paper]{iopart}

\newcommand{\eq}[1]{\begin{equation}#1\end{equation}}
\newcommand{\dd}{\mathrm{d}}
\newcommand{\ee}{\mathrm{e}}

\expandafter\let\csname equation*\endcsname\relax
\expandafter\let\csname endequation*\endcsname\relax 
\usepackage{amsmath}
\usepackage{iopams}
\usepackage{graphics}
\usepackage{graphicx}
\usepackage{psfrag}

\begin{document}

\title{Analytical results for the entanglement Hamiltonian of a free-fermion chain}
\author{ Viktor Eisler$^{1,2}$ and Ingo Peschel$^3$}
\address{
$^1$Institut f\"ur Theoretische Physik, Technische Universit\"at Graz, Petersgasse 16,
A-8010 Graz, Austria\\
$^2$MTA-ELTE Theoretical Physics Research Group, E\"otv\"os Lor\'and University,
P\'azm\'any P\'eter s\'et\'any 1/a, H-1117 Budapest, Hungary\\
$^3$Fachbereich Physik, Freie Universit\"at Berlin, Arnimallee 14, D-14195 Berlin, Germany
}

\begin{abstract}
We study the ground-state entanglement Hamiltonian for an interval of $N$ sites in a 
free-fermion chain with arbitrary filling. By relating it to a commuting operator, we find
explicit expressions for its matrix elements in the large-$N$ limit. The results agree
with numerical calculations and show that deviations from the conformal prediction persist
even for large systems.
\end{abstract}

\begin{center}
{\it Dedicated to John Cardy on the occasion of his 70th birthday.}
\end{center}

\maketitle

\section{Introduction}

In entanglement studies \cite{CCD09,Laflorencie16}, one divides a system into two parts and asks 
how these are coupled in the quantum state under consideration. This information is contained 
in the reduced density matrix for one of the subsystems which can be written in the form
$\rho=\exp(-\mathcal{H})/Z$ with an operator $\mathcal{H}$ now commonly called the 
entanglement Hamiltonian \cite{Li/Haldane08}. Its eigenvalue spectrum determines the degree 
of the entanglement and thus is of primary importance. However, also its structure is 
interesting since it shows universal features which depend on the way in which the system 
is divided.

For non-critical integrable chains divided into two halves, the relation of $\rho$ to corner 
transfer matrices leads to an operator $\mathcal{H}$ with terms which increase linearly
away from the boundary between the subsystems, see \cite{Peschel12}. This holds for spin chains 
\cite{Peschel/Kaulke/Legeza99} as well as for coupled oscillators 
\cite{Peschel/Chung99} and is also found in the continuum \cite{Gaite01}.
For critical quantum chains, conformal field theory gives results also for other cases,
as discussed most recently by Cardy and Tonni \cite{Cardy/Tonni16}. Thus, for a subsystem of
length $L$ in an infinite chain, one finds 
\begin{equation}
\mathcal{H} = 2\pi L \int_0^L dx \; \frac{x}{L}\left(1-\frac{x}{L}\right)\;T_{00}
\label{conf_ham}
\end{equation}
where $T_{00}$ is the energy density in the physical Hamiltonian \cite{Casini/Huerta/Myers11,Wong_etal13,
Wen_etal16, Cardy/Tonni16}. The parabolic weight factor becomes linear near the ends and can be viewed
as the combined effect of the two boundaries.

On the lattice, such a form was found in numerical studies of a fermionic hopping model 
equivalent to an XX chain \cite{review09} and in an XXZ chain for the value $\Delta=1/2$, 
when the ground state is particularly simple \cite{Nienhuis/Campostrini/Calabrese09}. However, 
the nearest-neighbour terms in $\mathcal{H}$ were not exactly parabolic and there were couplings
to more distant neighbours, albeit of smaller magnitude. Due to the small subsystems, no clear 
answer could be given for the asymptotic behaviour.

In this paper, we look once more at the free-fermion model and settle the problem for this
system. Starting from the insight obtained by numerical calculations, we develop an analytical
treatment in the large-$N$ limit. It is based on a paper by Slepian who studied the matrix 
problem underlying the determination of $\mathcal{H}$ already in the seventies in the context 
of time- and band-limited signals \cite{Slepian78,Eisler/Peschel13}. The strategy is to express 
$\mathcal{H}$ in terms of 
a closely related tridiagonal matrix $T$ and powers of it. For a half-filled system, this leads 
to explicit formulae for all the hopping matrix elements in terms of generalized hypergeometric 
functions. For other fillings, the coefficients in the resulting series are more complicated
and the summations have to be done numerically. In this way we confirm that the deviations
mentioned above do persist in the large-$N$ limit. In particular, the hopping to more distant 
neighbours for half filling shows a parabolic law raised to the power $r$ near the 
boundaries, where $r$ is the range. In the bulk, there is an additional enhancement and  
the maximal amplitude decreases with $r$ as $r^{-3}$. The analytical expressions are checked 
against numerical results obtained directly for large subsystem sizes. This requires extreme 
accuracy in the diagonalization of the correlation matrix from which $\mathcal{H}$ follows 
(roughly $N$ digits for $N$ sites) and such calculations have been done so far only in one very 
recent study where the continuum version of the present problem was addressed \cite{Arias_etal16}. 

In the following Section 2 we formulate the problem and give the basic expressions. Then, in 
Section 3, we present numerical results for the single-particle spectrum and the matrix elements of  
$\mathcal{H}$ for half filling. In Section 4 we recall some asymptotic results of Slepian and 
present the application to our problem with the resulting formulae for a half-filled system. In 
Section 5, the more complicated case of arbitrary filling is discussed and in Section 6 we conclude 
with a discussion. Results for the extremal eigenvalues and some mathematical details are collected 
in two appendices.

\section{Setting and basic formulae}

We consider the ground state of an infinite chain with the Hamiltonian 
\begin{equation}
\mathcal{\hat H} = -\frac{1}{2} \sum_n \,t\,(c^{\dag}_n c_{n+1} + 
                   c^{\dag}_{n+1} c_{n})+\sum_n\,d \, c^{\dag}_n c_n
\label{ham}
\end{equation}
where we take the hopping $t=1$ and use the site energy $d$ to adjust the ground-state filling.
With the single-particle energies $\omega_q=-\cos q+d$, the Fermi momentum is given by  $\cos q_F = d$.
All entanglement properties are determined by the correlation matrix 
$C_{m,n}=\langle c^{\dag}_m c_{n} \rangle$ in the ground state
\begin{equation}
C_{m,n} = \int_{-q_F}^{q_F} \frac{dq}{2\pi} \, e^{-iq(m-n)} =  
           \frac{\sin q_F(m-n)}{\pi (m-n)} \,.
\label{corr_latt1}
\end{equation}
In the following, we always consider a subsystem of $N$ consecutive sites denoted by $i$ or $j$. 
Since the ground state is a Slater determinant, the entanglement Hamiltonian $\mathcal{H}$ has 
again free-fermion form \cite{Peschel03}
\begin{equation}
\mathcal{H}=  \sum_{i,j=1}^N \, H_{i,j} c^{\dag}_i c_j
\label{ent_ham}
\end{equation}
and the matrix $H$ is given in terms of the eigenfunctions $\phi_k(i)$ and 
eigenvalues $\zeta_k$ of the correlation matrix $C_{i,j}$ restricted to the subsystem
\begin{equation}
H_{i,j}= \sum_{k=1}^N\,\phi_k(i)\; \varepsilon_k\; \phi_k(j)\
\label{spectral_H}
\end{equation}
where 
\begin{equation}
\varepsilon_k = \ln \frac {1-\zeta_k}{\zeta_k} \,.
\label{epsilon}
\end{equation}
\vspace{1mm}

An important (and intriguing) feature is that the restricted correlation matrix $C$ and thus 
the matrix $H$ commute with a tridiagonal matrix $T$ of the form 
\begin{equation}
   T =  
    \left(  \begin{array} {ccccc}
      d_1 & t_1   &  &   & \\  
       t_1 & d_2 & t_2  &  &\\
       & t_2 & d_3 & t_3  &  \\
        & & \ddots & \ddots & \\
        & & & t_{N-1} & d_N
  \end{array}  \right)   
  \label{tridiagonal1}
 \end{equation}
with elements 
 \begin{equation}
  t_i = \frac{i}{N} \left(1-\frac{i}{N}\right),\qquad
  d_i = -2\cos{q_F}\;\frac{2i-1}{2N}\;\left(1-\frac{2i-1}{2N}\right) .
  \label{tridiagonal2} 
 \end{equation}
This matrix can already be found in Slepian's paper \cite{Slepian78} (with diagonal terms differing 
from (\ref{tridiagonal2}) by an additive constant) and was later rediscovered in \cite{Peschel04}.
Its eigenvalues will be denoted by $\lambda_k$ and they will play a central role in the analytical 
treatment of the problem.

The matrices $C$, $H$ and $T$ have common eigenvectors and a common particle-hole symmetry. 
If one changes $q_F$ into $\pi-q_F$, they transform into $J(1-C)J,-JHJ$ and $-JTJ$, respectively, 
where $J$ is the diagonal matrix $J_{i,j}=(-1)^i\delta_{i,j}$. For the eigenvalues, when ordered 
according to magnitude, this means  
\begin{equation}
\varepsilon_k(q_F)=-\varepsilon_{N+1-k}(\pi-q_F), \quad \quad 
    \lambda_k(q_F)=-\lambda_{N+1-k}(\pi-q_F)
\label{symmetry1}
\end{equation}
and the corresponding eigenvectors satisfy 
\begin{equation}
 \phi_k(q_F,i)=(-1)^i\;\phi_{N+1-k}(\pi-q_F,i) \, .
\label{symmetry2}
\end{equation}
These relations will be used later in the construction of $H$. Pictorially speaking,
the $\varepsilon_k$ are mostly positive for small filling (since then most of the $\zeta_k$ 
which are occupation numbers lie near zero) and mostly negative for large filling. At half
filling, the spectrum is symmetric.

Finally, if one writes in analogy to (\ref{ent_ham})
\begin{equation}
\mathcal{T}=  \sum_{i,j=1}^N \, T_{i,j} c^{\dag}_i c_j
\label{T_ham}
\end{equation}
the operator $\mathcal{T}$ has, up to a factor of $-1/2$, \emph{exactly} the same form as the 
physical Hamiltonian (\ref{ham}) 
but with nearest-neighbour hopping and site energies which are both multiplied by essentially
the same parabolic factor increasing from the boundaries of the subsystem towards the middle. Thus 
it has the structure of the entanglement Hamiltonian (\ref{conf_ham}) as found in 
conformal field theory. The actual entanglement Hamiltonian $\mathcal{H}$ will turn out 
to be similar but not identical.

\section{Half filling: Numerical results}

\begin{figure}[thb]
\centering
\includegraphics[scale=.6]{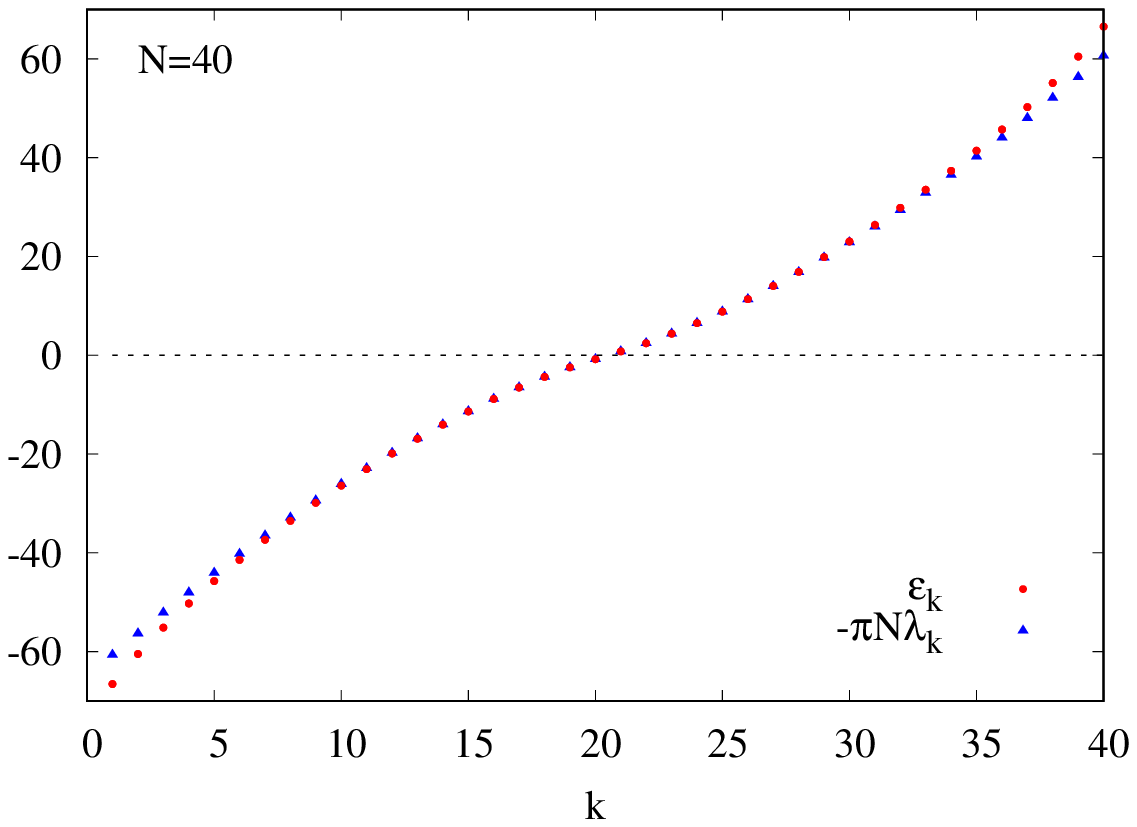}
\includegraphics[scale=.6]{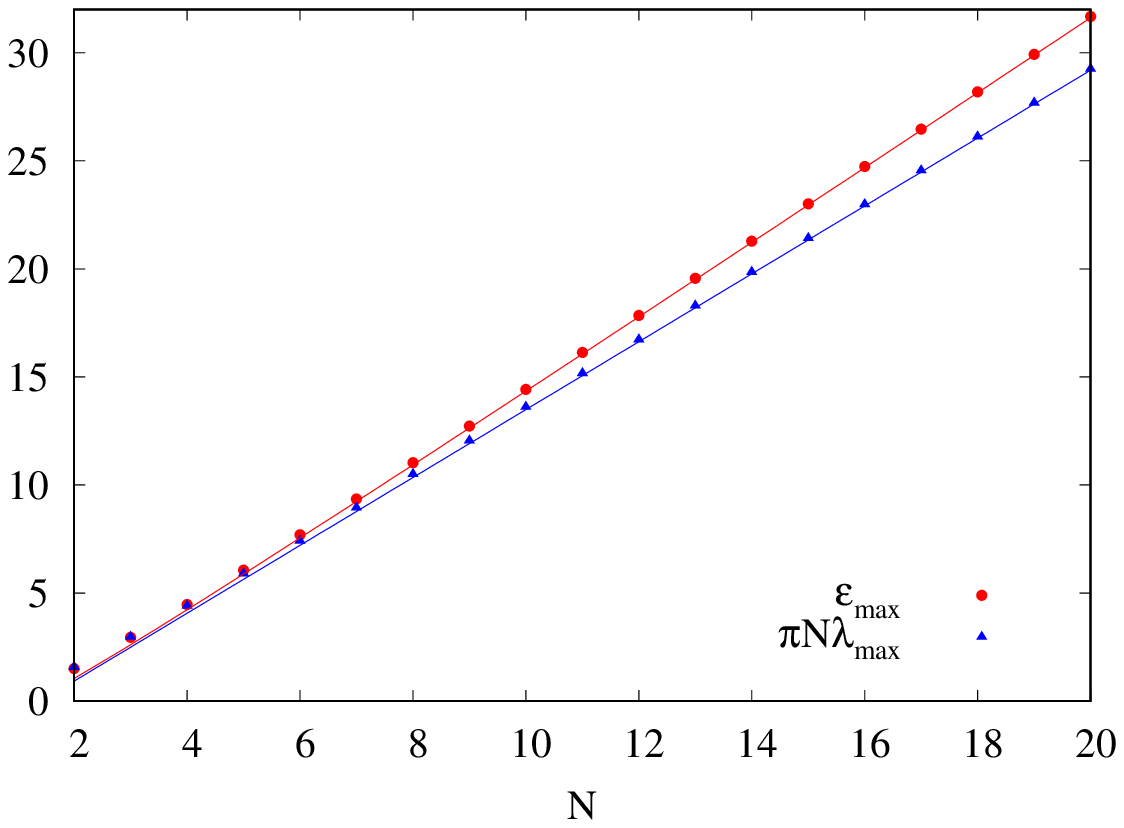}
\caption{Left: Eigenvalues $\varepsilon_k$ and $-\pi N \lambda_k$ for a half filled chain and $N=40$ 
               sites.
         Right: Maximal eigenvalues as functions of the size $N$. The lines are the 
                asymptotic formulae of Appendix A.}
\label{eps_lam_pi2}
\end{figure}

We present here the basic properties of $H$ and $T$ as obtained from numerical diagonalization 
for half filling, $q_F=\pi/2$.
It was noted already in \cite{Peschel04} that in this case the matrices $H$ and $\pi NT$  
have the same low-lying eigenvalues. This is shown in Fig. \ref{eps_lam_pi2} on the left 
where $\varepsilon_k$ and $-\pi N \lambda_k$ are plotted for a subsystem of $N=40$ sites. (The 
minus sign was left out in \cite{Peschel04}.) One sees that the small eigenvalues are practically 
indistinguishable on the chosen scale. Only for the large ones there is a small difference. 
Note that the largest $\varepsilon_k$ are about $\pm 66$ which means that the corresponding 
$\zeta_k$ differ from 0 or 1 by only $e^{-66} \simeq 10^{-29}$. Thus one needs at least 30 digits 
in the diagonalization of $C$ in order to find these $\varepsilon_k$ reliably. The actual 
calculation with \emph{Mathematica} used 40 digits.

The dependence of the maximal values $\varepsilon_{max}$ and $\pi N \lambda_{max}$ on the 
subsystem size $N$ is shown on the right of Fig. \ref{eps_lam_pi2}. Basically, both quantities
increase linearly. However, the slope is not the same and therefore their difference also
increases. This scaling with $N$ is also found analytically and will be discussed further in 
Section 4. The figure shows the excellent agreement between the numerics and the asymptotic
results already for small $N$. The scaling is not restricted to the maximal eigenvalues and 
suggests to work with the intensive quantities $H/N$ and $T$.

%
%

In Fig. \ref{elements1} the leading matrix elements in $-H/N$, calculated from (\ref{spectral_H}), are
displayed. Shown are the amplitudes for hopping to first, third and fifth neighbours as functions
of the position, again for $N=40$ sites. Hopping to second, fourth, etc. neighbours and diagonal
terms are strictly zero due to (\ref{symmetry1}),(\ref{symmetry2}). The figure resembles the one in 
\cite{review09} for $N=16$, but the scale factors to make the longer-range hopping visible are smaller 
here, i.e. these terms are relatively larger. Shown for comparison is also the nearest-neighbour hopping 
in $\pi T$ with its strictly parabolic form. Since the two matrices have the same eigenfunctions, all 
differences in their matrix elements are tied directly to the differences in the spectra. Although these
were seen to be relatively small, they occur for those eigenvalues which enter with the largest weight 
into the spectral representation (\ref{spectral_H}) and thus produce significant effects. In the next 
section, we will derive analytical formulae for the shown quantities.

%
\begin{figure}[thb]
\centering
\includegraphics[scale=.7]{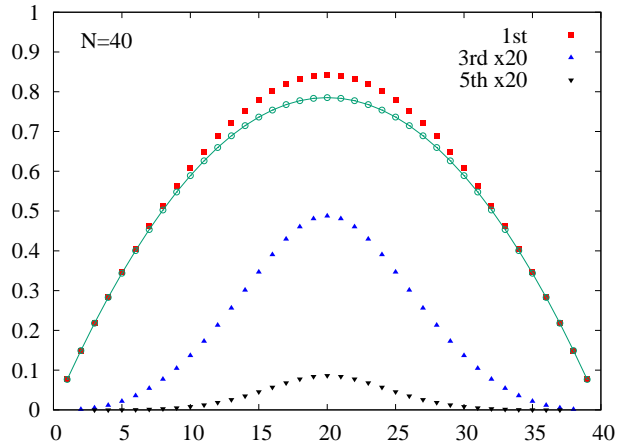}
\caption{Hopping matrix elements in $-H/N$ and nearest-neighbour hopping in $\pi T$ (green) as 
         functions of the position for $N=40$ sites.}
\label{elements1}
\end{figure}
%

\section{Half filling: Analytical treatment}

The half-filled case is particularly simple due to the invariance
of $H$ under a particle-hole transformation.
Indeed, using the symmetry properties \eref{symmetry1} and \eref{symmetry2}
of the spectrum and the eigenvectors, one can rewrite the entanglement
Hamiltonian as
\eq{
H_{i,j}= (1-(-1)^{i+j}) \sum_{k=1}^{N/2} \varepsilon_k \phi_k(i) \phi_k(j)
\label{Hij}}
where the sum involves only the negative branch of the spectrum.
The matrix elements $T_{i,j}$ have an analogous expression, with $\varepsilon_k$
replaced by the positive $\lambda_k$. As remarked in the previous section,
the operator $H$ has extensive character. It is thus useful to define
its density which, in the thermodynamic limit, is given by
\eq{
h_{i,j}=-\lim_{N \to \infty}\frac{H_{i,j}}{N} \, .
\label{hij}}

The main idea of the following calculation is to relate the spectra $-\varepsilon_k/N$
and $\lambda_k$ for $N \to \infty$ in terms of a series expansion.
From Fig. \ref{eps_lam_pi2} it is obvious, that to leading order the
relation is linear with a coefficient $\pi$. However, the behaviour
for larger eigenvalues hints at the presence of higher order
terms. To obtain the series analytically, we shall use the results of
\cite{Slepian78} where the following uniform asymptotic limit was found
\eq{
-\lim_{N \to \infty} \frac{\varepsilon_k}{N} = \eta(\kappa), \qquad
\kappa=\frac{k}{N} \quad \mathrm{fixed} \, .
}
The corresponding $\zeta_k$ are exponentially close to one,
$1-\zeta_k \sim \ee^{-N\eta(\kappa)}$, which is valid up to the centre
of the spectrum, $0< \kappa < 1/2$, in the scaling limit $k,N \gg 1$
with the ratio $\kappa = k/N$ kept fixed. The scaled eigenvalues are then
given by the integral$^{\footnotemark[1]}$\footnotetext{see Eqs. (43),(47) and (59) in \cite{Slepian78}.
Our $\zeta_k$ is called $\lambda_k$ there, our $\eta(\kappa)$ corresponds to
$L_3$ and our $q_F$ to the quantity $2\pi W$.}
\eq{
\eta(\kappa) =
\int_{0}^{B}  \sqrt{\frac{B-\xi}{\xi(1-\xi^2)}} \dd \xi
\label{epsint}}
where the dependence on the spectral parameter $\kappa$ is implicit
in the parameter $B=B(\kappa)$. The inverse of this function is given by
a similar integral
\eq{
\kappa(B) = \frac{1}{\pi}\int_B^1\sqrt{\frac{\xi-B}{\xi(1-\xi^2)}} \dd \xi
\label{kappa}}
where the limits of integration are changed. In particular,
the maximal eigenvalue $\kappa=0$ corresponds to $B=1$, 
where \eref{epsint} yields $\eta(0)=2 \,\mathrm{artanh\,} (1/\sqrt{2}) \simeq 1.7627$.
The subleading terms in the asymptotics of the extremal eigenvalues
can also be determined with the results given in Appendix A.
On the other hand, the centre of the spectrum, $\kappa=1/2$,
corresponds to $B=0$ and thus trivially $\eta(1/2)=0$.
In fact, these low-lying eigenvalues obey a different type of asymptotics,
scaling as $1/\ln N$ \cite{Slepian78,Peschel04}.

In a similar fashion, the eigenvalues $\lambda_k$ of $T$ can be analyzed and,
in the same uniform asymptotic limit, one finds$^{\footnotemark}$
\footnotetext{see Eqn. (133) in \cite{Slepian78}. The quantity $\theta_k$ there is
$ N^2(\lambda_k/2+ 1/4 \cos{q_F})$ due to the different definition 
of the tridiagonal matrix.} 
\eq{
\lim_{N \to \infty} \lambda_k = \lambda(\kappa)=\frac{B}{2}, \qquad
\kappa=\frac{k}{N} \quad \mathrm{fixed} \, .
\label{lamk}}
Thus the scaling limit $\lambda(\kappa)$ of the eigenvalues is,
up to a factor of two, simply given by the parameter $B$. Therefore,
instead of expressing $\eta(\kappa)$ and $\lambda(\kappa)$
in terms of the spectral parameter $\kappa$, one can find a
direct relation $\eta(\lambda)$ by expanding the integral \eref{epsint} 
in terms of $B$. Using the Taylor series
\eq{
\frac{1}{\sqrt{1-\xi^2}} = 
\sum_{m=0}^{\infty} \alpha_m \xi^{2m}, \qquad
\alpha_m = \frac{1}{\sqrt{\pi}} \frac{\Gamma(m+1/2)}{\Gamma(m+1)}
\label{taylor}}
one rewrites \eref{epsint} as an infinite sum of integrals
\eq{
\eta = \sum_{m=0}^{\infty} \alpha_m I_m \, .
\label{intm}}
These integrals can be evaluated explicitly and yield
\eq{
I_m = \int_{0}^{B} \sqrt{B-\xi}\; \xi^{2m-1/2}\, \dd \xi=
\beta_m \left(\frac{B}{2}\right)^{2m+1}, \qquad
\beta_m = \sqrt{\pi}\; 2^{2m}  \frac{\Gamma(2m+1/2)}{\Gamma(2m+2)}.
\label{intm2}}
Hence, using Eq. \eref{lamk}, one has the series expansion
\eq{
\eta = \sum_{m=0}^{\infty}
\alpha_m \beta_m \lambda^{2m+1} \, .
\label{epslam}}
As expected, the coefficient of the linear ($m=0$) term is $\alpha_0 \beta_0 = \pi$. 

Our next goal is to rewrite the relation between the eigenvalues
in terms of the corresponding matrices. This can be done by substituting
\eref{epslam} into \eref{Hij} and \eref{hij} and taking the sum over the
eigenvalues in the proper scaling limit. Since only odd powers of $\lambda$
appear in \eref{epslam}, the symmetry used in \eref{Hij} holds for 
all terms and the relation can immediately be lifted to the matrix level
\eq{
h = \sum_{m=0}^{\infty}
\alpha_m \beta_m T^{2m+1} \, .
\label{ht}}
The above exact series representation of $h$ already explains the two important
qualitative features found numerically in the previous section, namely the deviation
from the parabolic profile in the nearest-neighbour hopping, and the appearance
of longer-range hopping. Indeed, the $n$-th power of the matrix $T$ generates
hopping terms up to the $n$-th neighbour.

To evaluate the matrix elements $h_{i,i+r}$ in the scaling limit, one needs a closed-form
expression for the corresponding matrix elements of the powers of $T$
\eq{
(T^{2m+1})_{i,i+r} = \sum_{l_1,\dots,\l_{2m}}
T_{i,l_1} T_{l_1,l_2} \dots
T_{l_{2m},i+r} \, .
\label{tmult}}
The summation indices run over the values $l_j=1,\dots,N$ for each $j=1,\dots,2m$.
However, due to the tridiagonal structure of $T$ with the diagonal being identically
zero for half-filling, the only non-vanishing terms in the sum occur for indices
that satisfy the constraints
\eq{
| l_{j}- l_{j-1}| = 1, \qquad j=1,\dots,2m+1, \qquad l_0 = i, \qquad l_{2m+1}=i+r
\label{rw}}
where we defined extra indices $l_0 $ and $l_{2m+1}$ that are kept fixed.
These constraints define a simple random walk, with the position of the
walker given by $l_j$, and with the requirement that the walk starts at site
$i$ and arrives at $i+r$ in exactly $2m+1$ steps. Since the number of steps
is always odd, so must be the distance covered $r=2p+1$, and thus the only
non-vanishing matrix elements are $h_{i,i+2p+1}$, as already clear from
Eq. \eref{Hij}.

The expression in Eq. \eref{tmult} further simplifies if one takes a proper
continuum limit $i,N \to \infty$ with $i/N$ kept fixed. Indeed, considering the
definition of the matrix elements in \eref{tridiagonal2}, all the factors $t_{l_j}$
with $j=0,\dots, 2m$ appearing in \eref{tmult} can be chosen to be equal. 
Even though the range of sites visited during the random walk grows as
$|i-l_j| \le m-p$, for any \emph{finite} and \emph{fixed} values of $m$ and $p$
one has $t_{l_0} \simeq t_{l_1} \simeq \dots \simeq t_{l_{2m}}$ up to corrections
of the order $1/N$ that vanish in the continuum limit. In choosing the proper
scaling variable one should, however, take into account the reflection symmetry
on the lattice, i.e. $h_{i,i+2p+1}$ should be invariant under $i \to N-2p-i$.
The continuum limit of \eref{tmult} satisfying this constraint then reads
\eq{
(T^{2m+1})_{i,i+2p+1} = \binom{2m+1}{m-p}\, z^{2m+1}, \qquad
z = t_{i+p}=\frac{i+p}{N}\left(1-\frac{i+p}{N}\right)
\label{tcont}}
where the combinatorial factor is just the number of random walks
satisfying \eref{rw} with $r=2p+1$ and the quantity $i+p$ in $z$ is essentially 
the midpoint between initial and final site in the hopping.

Finally, the result \eref{tcont} must be inserted into \eref{ht}.
Interestingly, the infinite sum can be carried out and yields the
following closed-form expression
\vspace{0.2cm}
\eq{
h_{i,i+2p+1} =
\pi\frac{(2p-1)!!(4p-1)!!}{2^p \, p! (2p+1)!}\; z^{2p+1} \,
_{3}F_{2}\left(p+\frac{1}{4},p+\frac{1}{2},p+\frac{3}{4}; p+1,2p+2 ; (4z)^2\right)
\label{hpscf}
\vspace{0.2cm}}
where $(2p-1)!!=1\cdot 3\cdot 5...(2p-1)$ and $_{3}F_{2}$ is a generalized hypergeometric 
function. Its definition as well as the derivation of the result \eref{hpscf}
are summarized in Appendix B. In particular, for the
nearest-neighbour hopping one has
\eq{
h_{i,i+1} = \pi \, x(1-x) \, 
_{3}F_{2}\left(\frac{1}{4},\frac{1}{2},\frac{3}{4}; 1,2 ; \left[4x(1-x)\right]^2\right)
\label{hscf}}
with $x=i/N$. One thus finds that the parabolic profile is multiplied with
a generalized hypergeometric function which increases monotonously
from $1$ at $x=0$ up to its maximum of approximately $1.076$ in the middle
of the interval, $x=1/2$. Hence, despite the continuum limit involved, the
lattice result for the nearest-neighbour hopping is different from the CFT prediction,
with a relative deviation of approximately $8\%$ around the maximum.

\begin{figure}[htb]
\center
\includegraphics[width=0.49\columnwidth]{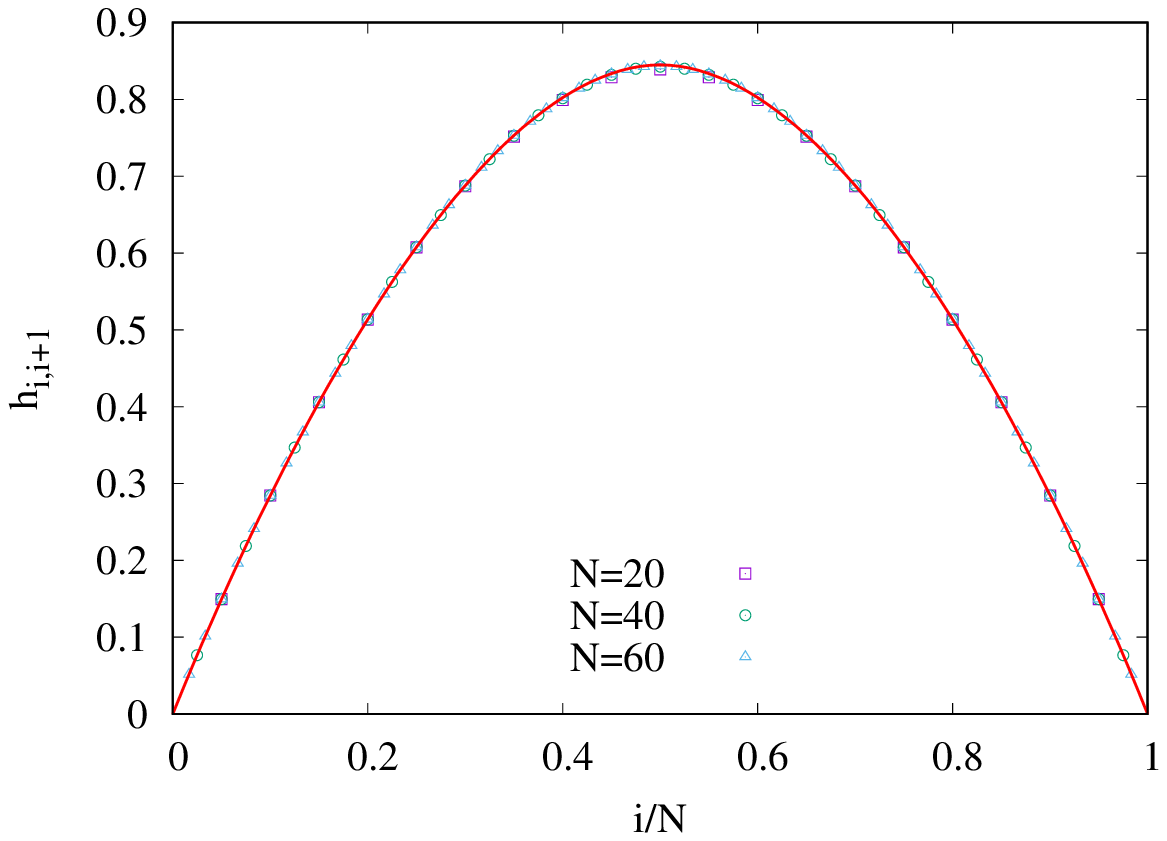}
\includegraphics[width=0.49\columnwidth]{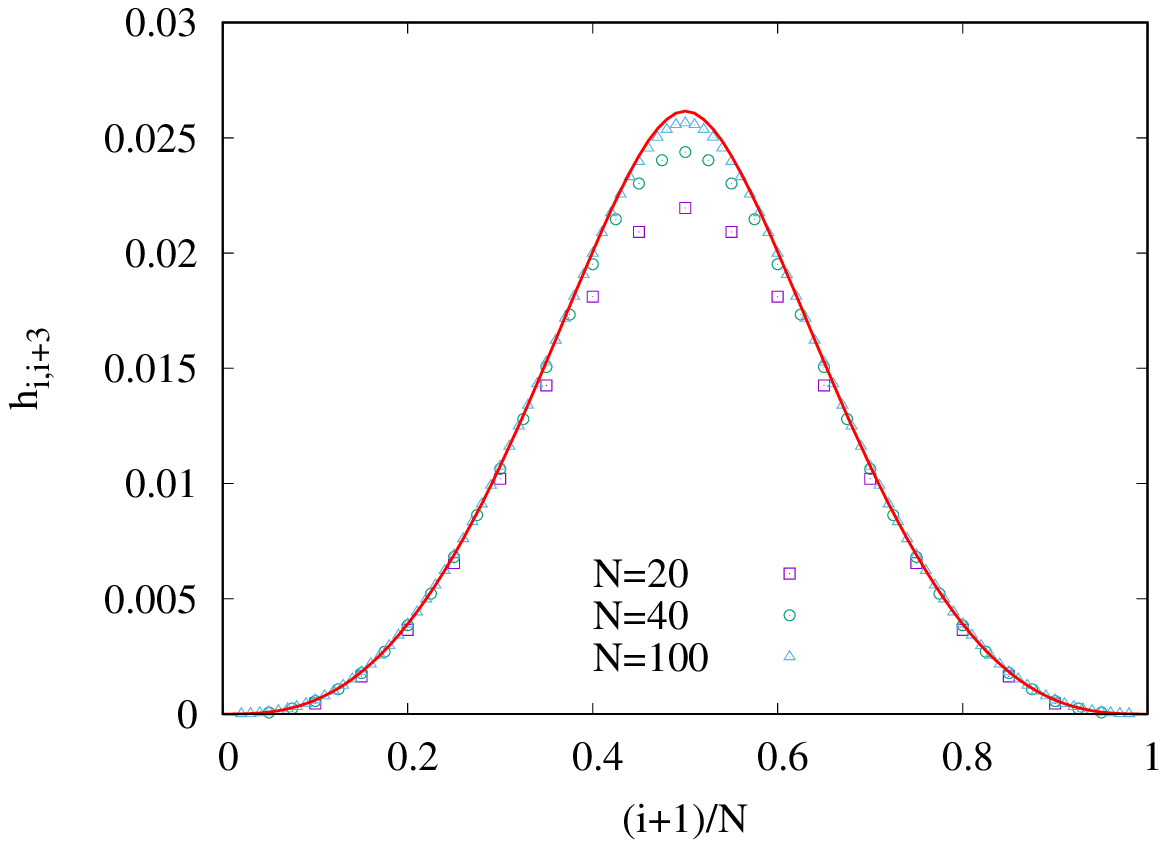}
\caption{Nearest-neighbour ($p=0$, left) and third-neighbour ($p=1$, right) hopping
in the scaled entanglement Hamiltonian $h$. The numerical results are shown by
symbols for various $N$, while the red solid line is the analytical result \eref{hpscf}.}
\label{fig:scf}
\end{figure}

The analytical result \eref{hpscf} for the scaled entanglement Hamiltonian
can be tested against direct lattice calculations as in Section 3 and for increasing 
values of $N$.
This is shown in Fig. \ref{fig:scf} for the nearest ($p=0$) and third-neighbour
($p=1$) hopping. While for $p=0$ the agreement is excellent already for relatively
small $N$, the case $p=1$ already requires to take $N=100$ to obtain
a good convergence. This can be attributed to the much smaller value of the matrix 
element. The finite-size corrections that have been neglected in $\eref{tcont}$ 
are then more pronounced.
Nevertheless, the data shows a perfect convergence towards the analytical
result.

%
%

\begin{figure}[htb]
\center
\includegraphics[width=0.49\columnwidth]{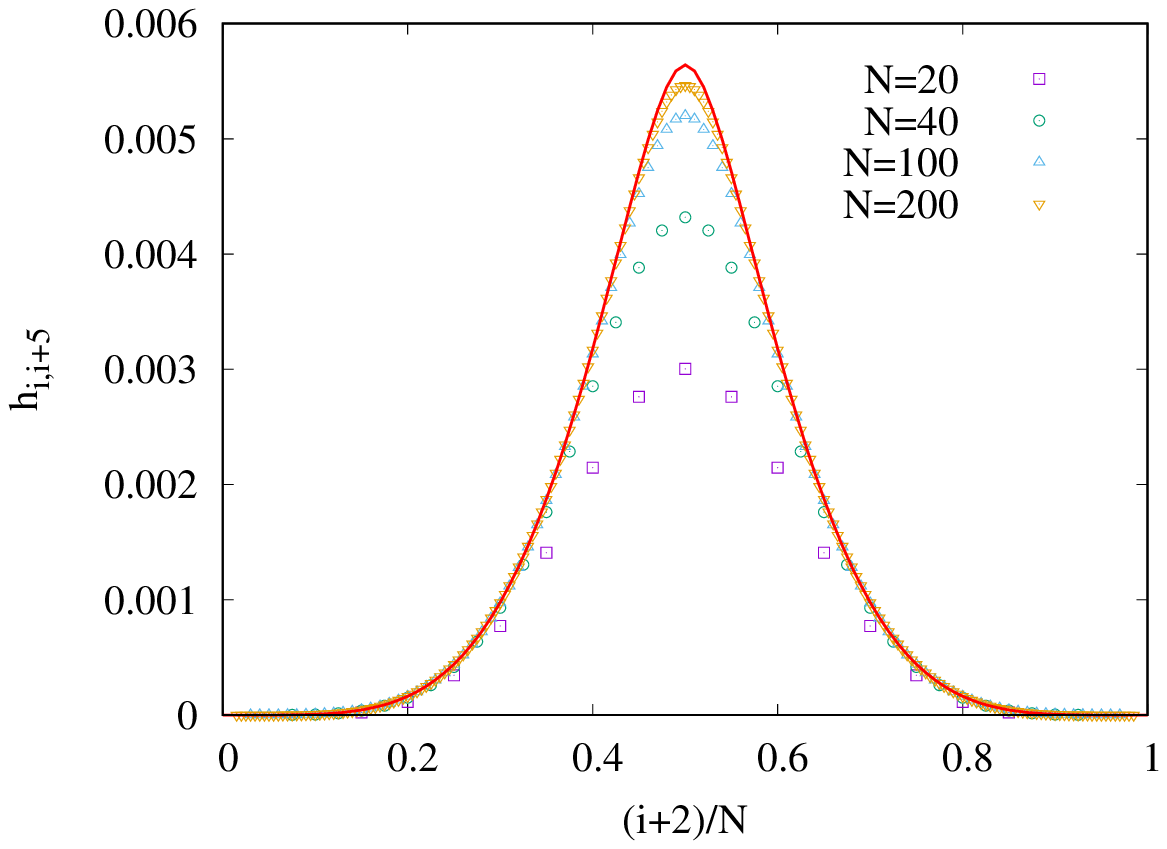}
\includegraphics[width=0.49\columnwidth]{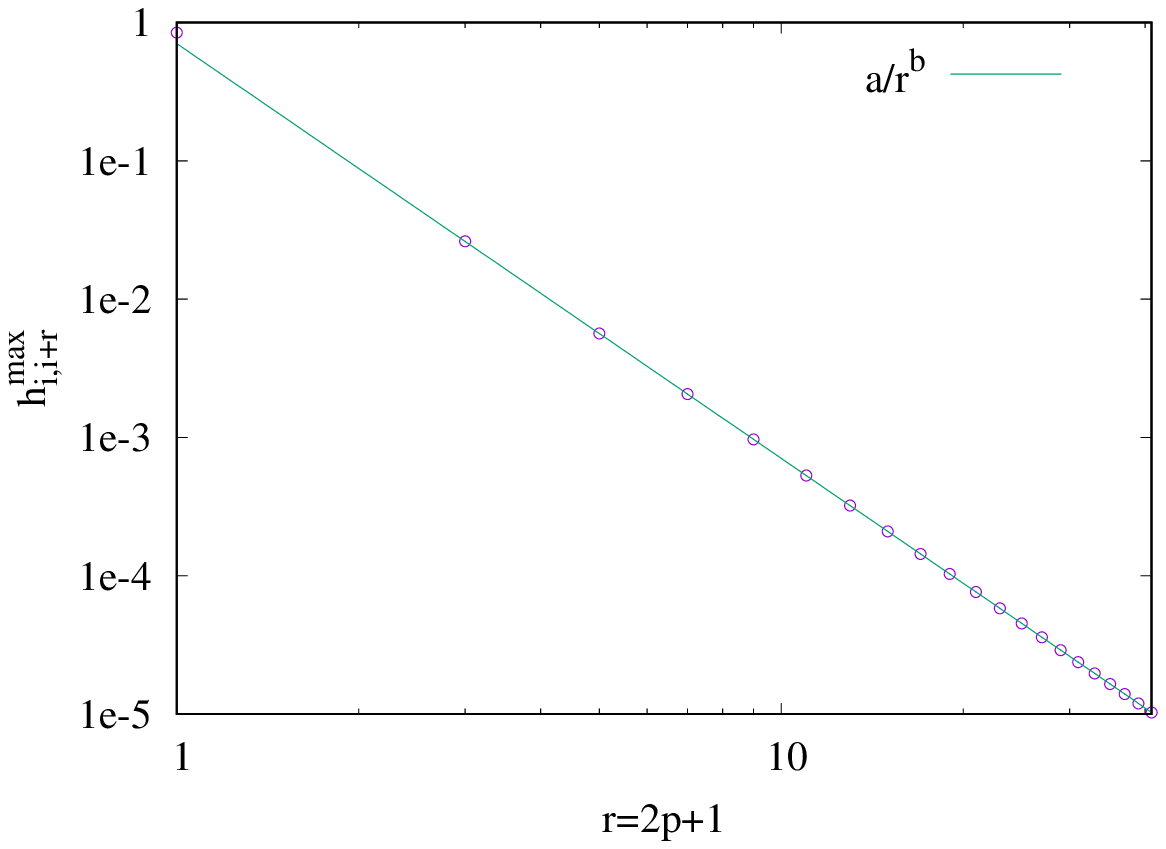}
\caption{Left: fifth-neigbour hopping in $h$, with numerical data (symbols)
compared to the analytical result (red solid line) of Eq. \eref{hpscf} with $p=2$.
Right: maxima (symbols) of the long-range hopping terms as a function of the
distance $r=2p+1$. The solid line is a power-law fit with $a=0.705$ and $b=2.999$.}
\label{fig:scf2}
\end{figure}

This trend continues also for larger $p$ values. The fifth-neighbour
hopping is shown on the left of Fig. \ref{fig:scf2}, and here the largest interval
size had to be taken as $N=200$ in order to reach a convincing agreement. Note that
the convergence is always the slowest around the peaks. In fact, it is
interesting to have a look at the decay of these maxima as a function of
the distance $r=2p+1$ in the hopping, i.e. evaluating \eref{hpscf}
at the argument $4z=1$. This is shown on the right of Fig. \ref{fig:scf2}
on a logarithmic scale. One can see a clear power-law decay with
a power $3$ which was obtained by fitting the data. Formally, it results from
an increase of $_{3}F_{2}$ with $r$ which is overcompensated by a decrease of the 
prefactor. While the drop from $r=1$ to $r=3$ is rather big, the following maxima
decrease less rapidly. Note that if one plotted the maxima for a \emph{fixed} value 
of $N$, the decay would be faster due to the slower convergence for higher $p$.

As to the hopping profiles, the curves are more and more 
concentrated in the middle of the system as $r$ becomes larger. Near the boundaries,
this simply reflects the factor $z^{2p+1}$ in \eref{hpscf} which originates from
the lowest power of $T$ leading to the corresponding matrix element.  
In the centre, the corrections contained in $_{3}F_{2}$ enhance the values and 
lead to a sharpening of the profile. 

%
%

\section{Arbitrary filling}

\begin{figure}[htb]
\center
\includegraphics[width=0.49\columnwidth]{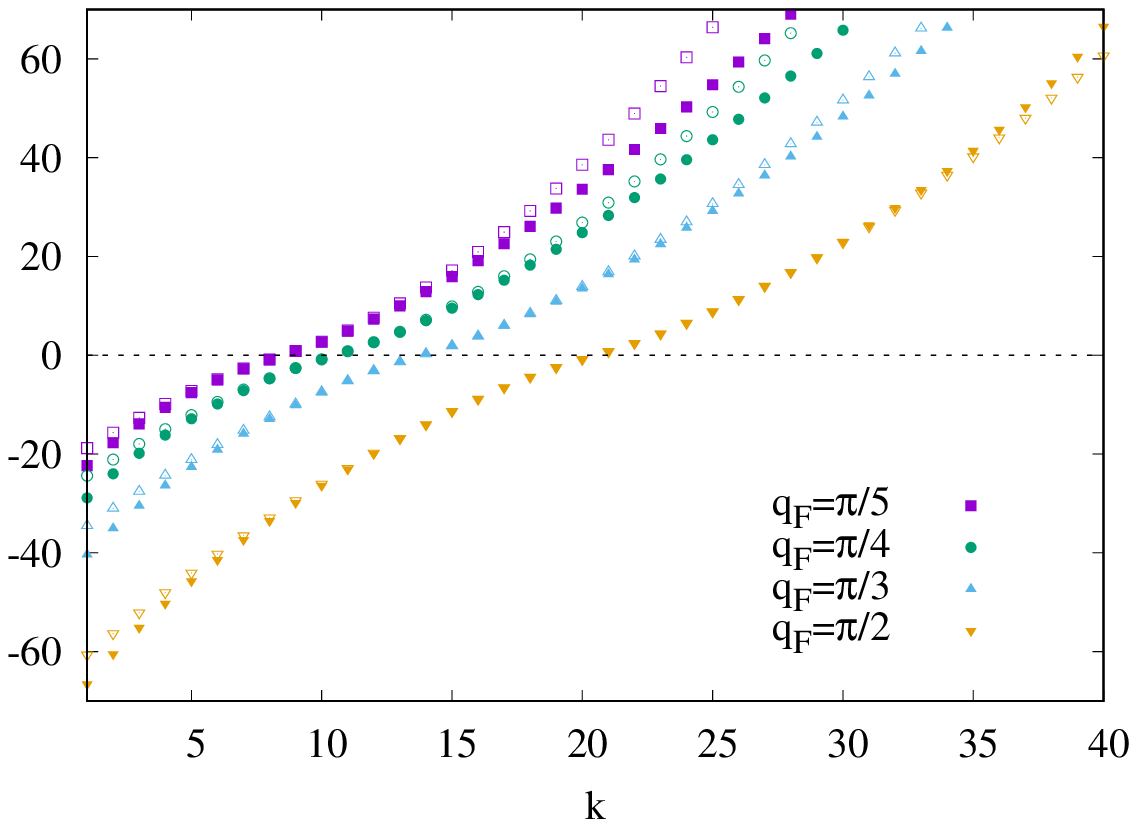}
\includegraphics[width=0.49\columnwidth]{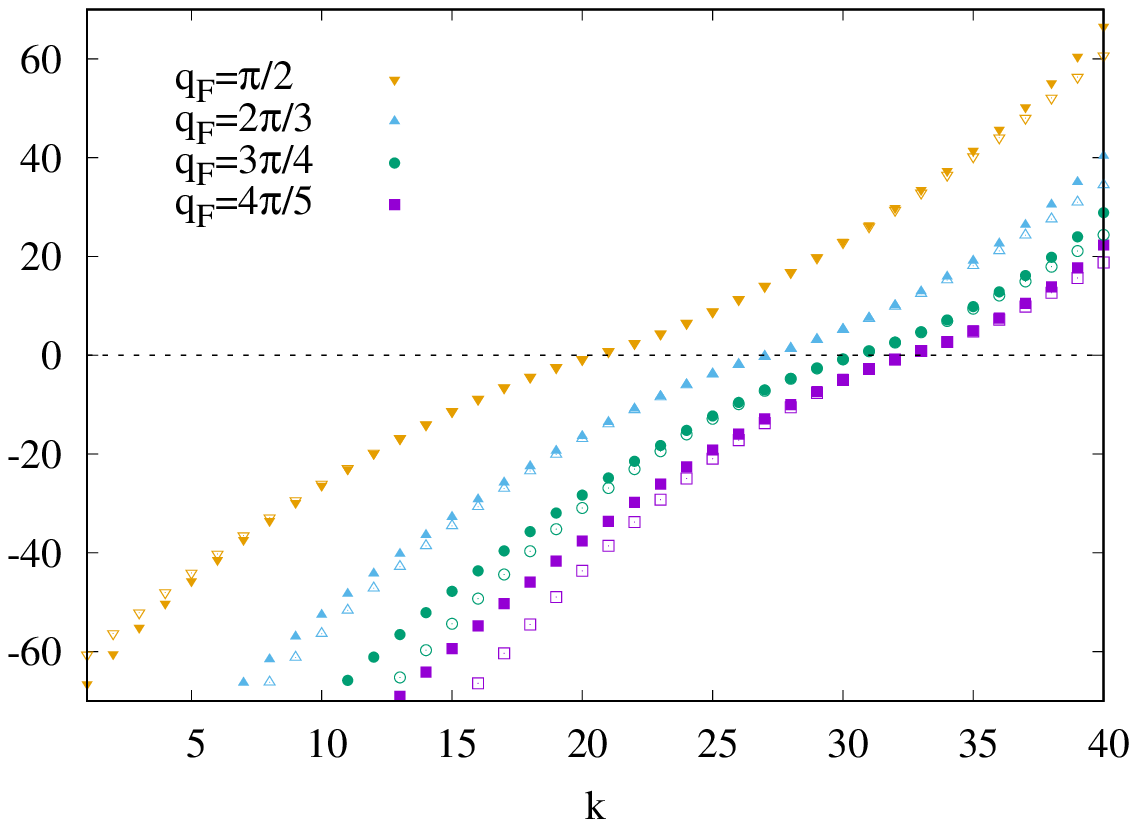}
\caption{Spectra for $N=40$ and different fillings.
 Left: $q_F \le \pi/2$. Right: $q_F \ge \pi/2$. Results for $q_F$ and $\pi-q_F$ are shown 
 in the same colours. Full symbols: $\varepsilon_k$, open symbols: $-\pi N \lambda_k/ \sin q_F$. 
 }
\label{eps_lam_gen}
\end{figure}

We proceed to the treatment of the general case, where the filling is given by
$q_F/\pi$ with arbitrary $0<q_F<\pi$. We first show the spectra for several values of $q_F$
in Fig. \ref{eps_lam_gen}. The symmetry stated in \eref{symmetry1} is clearly visible, even though
the largest eigenvalues fall outside the figure. The quantities $-\pi N \lambda_k$ are divided
here by an additional factor $\sin q_F$ which is motivated by the analytics below and seen to 
produce good agreement with the low-lying $\varepsilon_k$. However, the higher $\varepsilon_k$ now 
lie either fully above or fully below the $\lambda$-values which hints at even powers in a series 
expansion.
  
%
%
The idea of the analytical treatment is the same as before, but the formulas and technical 
details are more involved. We start again by
using the symmetries \eref{symmetry1} and \eref{symmetry2} to write
\eq{
H_{i,j} = \sum_{k=1}^{\bar N} \varepsilon_k \phi_k(i) \phi_k(j)
-(-1)^{i+j} \sum_{k=1}^{N-\bar N} \hat\varepsilon_k \hat\phi_k(i) \hat\phi_k(j)
\label{Hij2}}
where $\bar N \simeq N q_F/\pi$ is the number of $\varepsilon_k$
eigenvalues in the negative branch and the hat denotes quantities corresponding to 
$\pi-q_F$. An analogous formula holds for $T_{i,j}$, with the sums taken over the positive
branches of $\lambda_k$ and $\hat \lambda_k$ eigenvalues.
Introducing the density \eref{hij} as in the half-filled case, one has
the matrix representations
\eq{
h = h_- - J \hat h_- J \, , \qquad
T = T_+ - J \hat T_+ J \, .
\label{hmtm}}
The goal is again to relate the above operators in terms of a series
expansion. However, in contrast to the half-filled case, one has now
$h_- \ne \hat h_-$ and $T_+ \ne \hat T_+$ and thus the two contributions
have to be treated separately at first. Nevertheless, it is easy to see that
the two pieces in the decomposition \eref{hmtm} are orthogonal,
$h_- (J\hat h_- J) =0$ and similarly for $T_+$, which will be used later.

As before, the first step is to consider the uniform asymptotic limit of the
eigenvalues. Instead of \eref{epsint} and \eref{lamk}, one has now
\eq{
\eta(\kappa) = \int_{A}^{B} \sqrt{\frac{B-\xi}{(\xi-A)(1-\xi^2)}} \, \dd \xi, \qquad
\lambda(\kappa) = \frac{B-A}{2}, \qquad A=\cos q_F \, .
\label{epsint2}}
Analogously, the general form of the spectral parameter is given by
\eq{
\kappa(B) = \frac{1}{\pi}\int_B^1\sqrt{\frac{\xi-B}{(\xi-A)(1-\xi^2)}} \,\dd \xi
\label{kappa2}}
where the maximal eigenvalue, $\kappa=0$, corresponds again to $B=1$,
whereas setting $B=A$ yields $\kappa=q_F/\pi$, i.e. the endpoint of the
branches with $\eta(q_F/\pi)=\lambda(q_F/\pi)=0$.
For $B \simeq A$, \eref{epsint2} can be evaluated by setting $\xi^2 = A^2$ in the integrand
with the result  
\eq{
\eta =  \frac{\pi}{2} \frac{B-A}{\sqrt{1-A^2}} = \frac{\pi \lambda}{\sin q_F}
\label{epsint3}}
which was already used in plotting the numerical data. Note that the factor $\sin q_F$ can be seen
as the Fermi velocity in \eref{ham} for general filling.

To obtain the function $\eta(\lambda)$ in general, an expansion around $\xi=A$ is, however,
not useful, since the resulting series do not converge for all $A$. Instead, we insert again the Taylor 
series \eref{taylor} into the integral in \eref{epsint2}. This gives the infinite sum
\eref{intm} with the integrals
\eq{
I_m = \int_{A}^{B} \sqrt{\frac{B-\xi}{\xi-A}}\, \xi^{2m}  \dd \xi=
\sum_{n=0}^{2m+1} \beta_{m,n}
\left(\frac{B}{2}\right)^{2m+1-n} \left(\frac{A}{2}\right)^n
\label{intm3}}
where the coefficients read
\eq{
\beta_{m,n} = 2^{2m} \frac{\Gamma(2m-n+1/2)\Gamma(n+1/2)}{\Gamma(2m-n+2)\Gamma(n+1)} \, .
\label{betamn}}
Hence, $\eta$ is now given in terms of a double series in powers of
both parameters $A$ and $B$. Using the expression for $\lambda$ in
\eref{epsint2}, this can be written as
\eq{
\eta = 
\sum_{m=0}^{\infty}\sum_{n=0}^{2m+1}  \alpha_m \beta_{m,n}
\left(\lambda + \frac{A}{2}\right)^{2m+1-n} \left(\frac{A}{2}\right)^n .
\label{epslam2}}
For half filling, where $A=0$ and only the term
$n=0$ contributes, one recovers 
the previous formula \eref{epslam}. Furthermore, the inversion
of the filling, $q_F \to \pi-q_F$, corresponds to $A \to -A$, and
thus an analogous expression with the appropriate sign change
holds between $\hat \eta$ and $\hat \lambda$. The fact that $\eta$ vanishes
for $\lambda=0$ is not directly visible but hidden in the coefficients $\beta_{m,n}$,
of which $\beta_{m,2m+1}$ is negative.

Lifting \eref{epslam2} to the matrix level requires
some extra considerations. Although $\eta(\lambda)$
(respectively $\hat\eta(\hat\lambda)$) can straighforwardly be written
as a relation between the matrices $h_-$ and $T_+$
(respectively $\hat h_-$ and $\hat T_+$), this does not immediately
yield a relation between $h$ and $T$ as given in Eq. \eref{hmtm}.
Incorporating the matrix factors $J$ is no problem, but
all contributions involving the $T_+$ have to combine correctly.
Now, expanding in powers in \eref{epslam2}, one obtains terms of the form $T_+^{k}A^l$
where the sum $k+l=2m+1$ is always odd. In the expression for inverted filling, the terms 
with odd powers of $A$ (and thus even powers of $T_+$) will enter with changed sign.
Hence, in the sum for $h_- - J \hat h_- J$ one has only the combinations
\eq{
T_+^{2k+1}-(J\hat T_+ J)^{2k+1} \equiv T^{2k+1}\;, \qquad
T_+^{2k}+(J\hat T_+ J)^{2k} \equiv T^{2k}
\label{tpm}}
where the orthogonality of the two pieces was used. Therefore, the series
can indeed be rewritten in terms of $T$ only and one has
\eq{
h = \sum_{m=0}^{\infty} \sum_{n=0}^{2m+1}
\alpha_m \beta_{m,n}
\left(T+\frac{A}{2}\right)^{2m+1-n} \left(\frac{A}{2}\right)^n .
\label{ht2}}

In the final step, we generalize the calculation for the matrix elements to powers
of the shifted matrix $\tilde T = T + A/2$
\eq{
(\tilde T^{s})_{i,i+r} = \sum_{l_1,\dots,\l_{s}}
\tilde T_{i,l_1} \tilde T_{l_1,l_2} \dots
\tilde T_{l_{s},i+r} \, .
\label{tmult2}}
The power $s$ can now take both even and odd values and the summation indices $l_j$ 
must satisfy the following constraints
\eq{
| l_{j}- l_{j-1}| \le 1, \qquad j=1,\dots,s+1, \qquad l_0 = i, \qquad l_{s+1}=i+r \, .
\label{rw2}}
Contrary to the half-filled case, now $l_j-l_{j-1}=0$ is also allowed since
the diagonal elements of $\tilde T$ are nonvanishing. Therefore, Eq. \eref{rw2}
defines a random walk, where at each step $j$ the walker is allowed to
pause, such that the final site $i+r$ is reached from $i$
in exactly $s$ steps. Denoting by $\ell$ the number of stops,
one has in the continuum limit
\eq{
(\tilde T^{s})_{i,i+r} = \sum_{\substack{\ell=0  \\ s-r-\ell \,\, \mathrm{even}}}^{s-r}
\binom{s-\ell}{\frac{s-r-\ell}{2}} \binom{s}{\ell} \left(\frac{A}{2}-2Az \right)^\ell z^{s-\ell}
\label{tcont2}}
where the first binomial factor gives the number of walks
reaching the destination site at a distance $r$ in $s-\ell$ jumps, while the second
one gives the number of ways the $\ell$ stops can be chosen within $s$ steps.
Clearly, the number of jumps $s-\ell$ must have the same parity as the distance
$r$, which restricts the possible terms in the sum. Finally, the two factors at the end
of \eref{tcont2} are just the diagonal and offdiagonal matrix elements of $\tilde T$
in the continuum limit, raised to the power $\ell$ and $s-\ell$, respectively.
As for the half-filled case, the scaling variable $z$ must be
chosen such that the lattice reflection symmetry $i \to N+1-r-i$ is satisfied
for $h_{i,i+r}$, which yields
\eq{
z = \frac{2i+r-1}{2N}\left(1-\frac{2i+r-1}{2N}\right) .
\label{zr}}

One thus arrives at the final expression 
\eq{
h_{i,i+r} = \sum_{m=0}^{\infty} \sum_{n=0}^{2m+1}
\alpha_m \beta_{m,n}
\left(\tilde T^{2m+1-n}\right)_{i,i+r}
\left(\frac{A}{2}\right)^n
\label{ht3}}
where \eref{tcont2} with $s=2m+1-n$ has to be inserted for the matrix elements. 
Since this expression is significantly
more complicated than the one for the half-filled case, we refrain from
a further analysis. Instead, we simply evaluate \eref{ht3} numerically,
using a sufficiently large cutoff $M$ in the infinite sum. 

\begin{figure}[htb]
\center
\includegraphics[width=0.49\columnwidth]{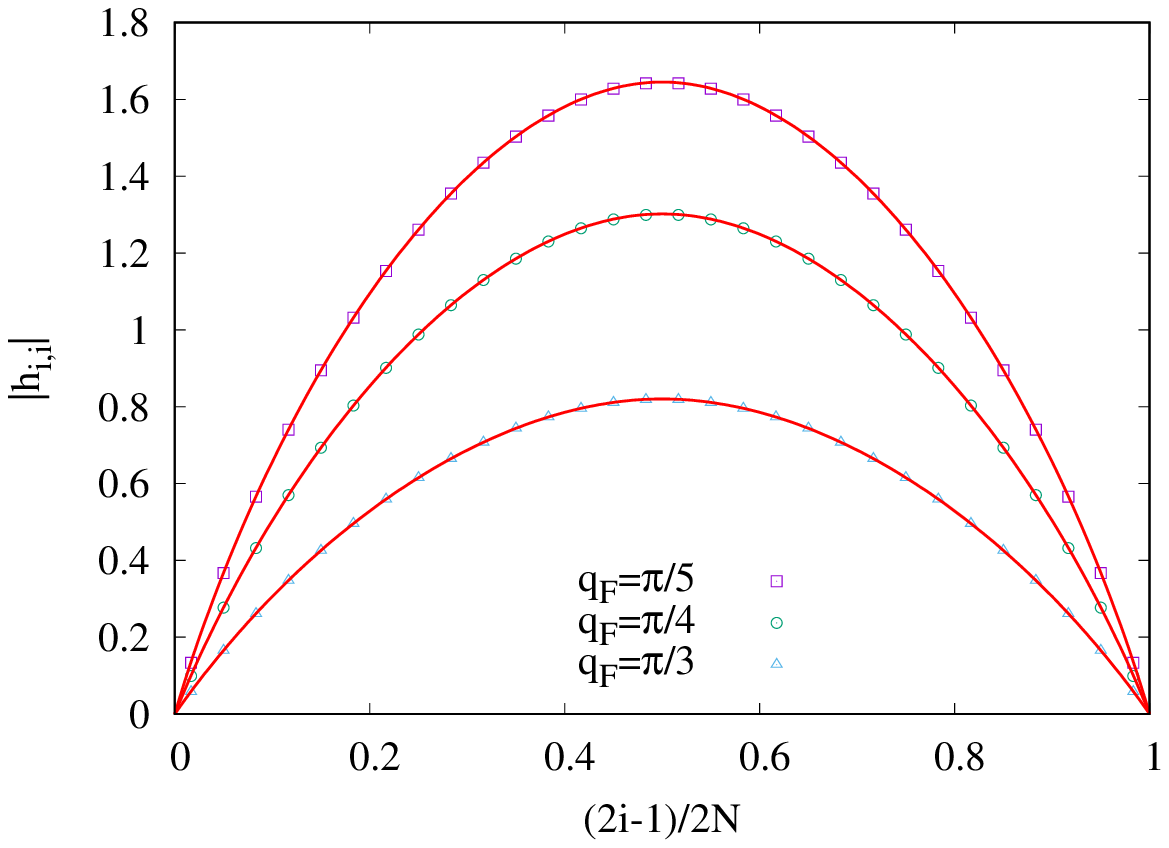}
\includegraphics[width=0.49\columnwidth]{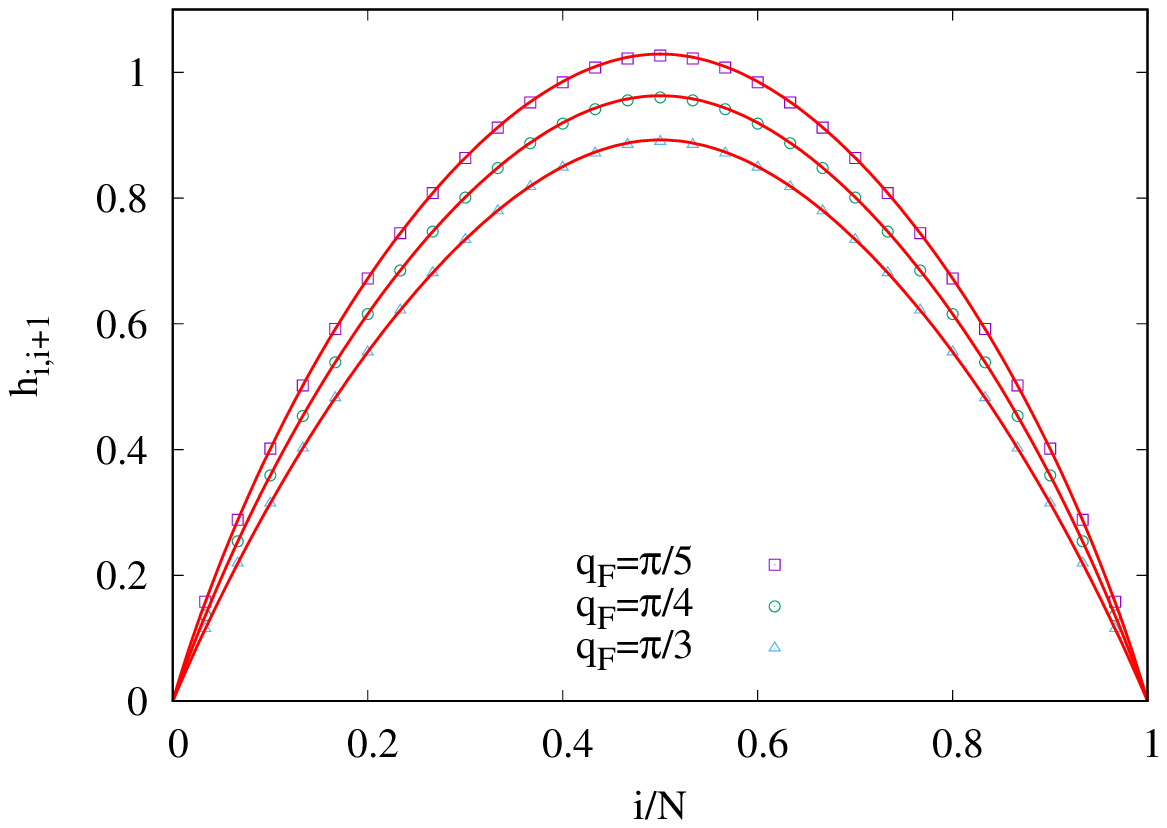}
\caption{Matrix elements of $h$ for various fillings.
The numerical data (symbols) for $N=40$ is compared to the analytical result
\eref{ht3} (lines), with a cutoff $M=30$ in the sum.
Left: diagonal terms. Right: nearest-neighbour hopping. Note the different vertical scales.}
\label{fig:scf3}
\end{figure}

The scaling functions 
obtained in this way are compared to the lattice results in
Fig. \ref{fig:scf3} for the diagonal ($r=0$, left) and nearest-neighbour
hopping terms ($r=1$, right). In both cases we have considered an
interval of length $N=40$ and the cutoff was set to $M=30$, yielding
an excellent agreement for all the different fillings shown. We have also
checked that decreasing the cutoff to $M=10$ has no visible effects
on the scaling functions, thus the convergence
of \eref{ht3} is very fast for the dominant matrix elements of $h$.
As the diagonal entries $T_{i,i}$ in \eref{tridiagonal2}, the matrix elements 
$h_{i,i}$ are negative for $q_F < \pi/2$ and we plotted their absolute value. 

%
%

\begin{figure}[htb]
\center
\includegraphics[width=0.49\columnwidth]{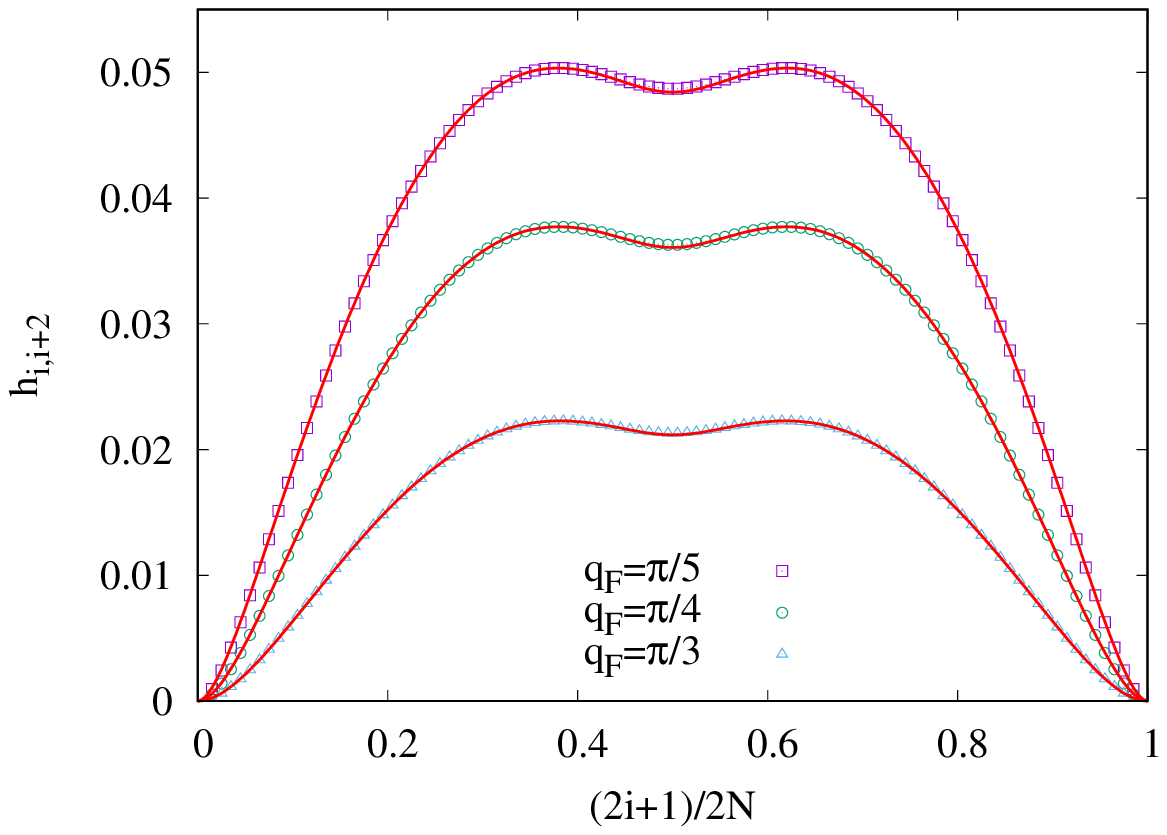}
\includegraphics[width=0.49\columnwidth]{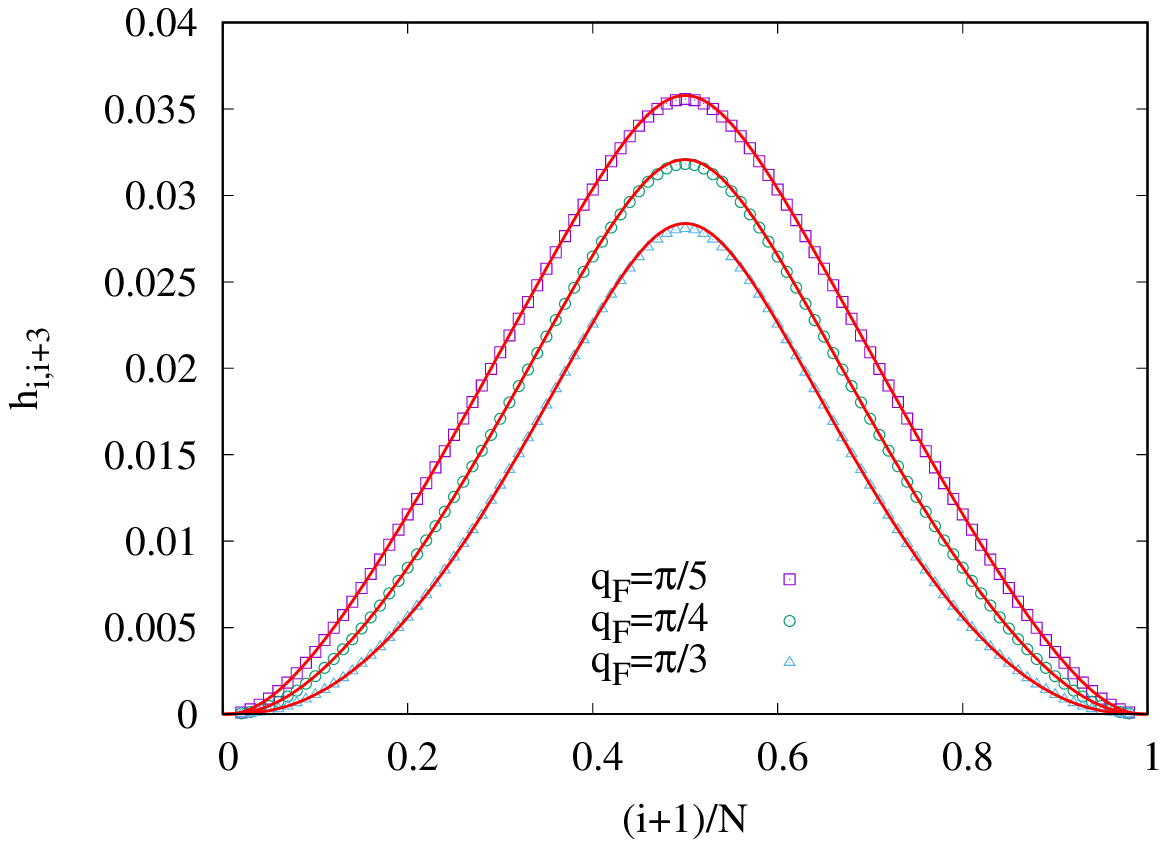}
\caption{Matrix elements of $h$ for various fillings.
The numerical data (symbols) for $N=100$ is compared to the analytical result
\eref{ht3} (lines), with a cutoff $M=50$ in the sum.
Left: second-neighbour hopping. Right: third-neighbour hopping.}
\label{fig:scf4}
\end{figure}

It is interesting to have a look also at the longer-range hopping terms,
shown in Fig. \ref{fig:scf4} for $r=2$ and $r=3$, which have again much
smaller amplitude. As for half-filling, this requires to take larger intervals
for a good agreement with the continuum limit result. We thus used
$N=100$ in Fig. \ref{fig:scf4} and increased also the cutoff to $M=50$,
leading again to a perfect match. Note that setting $M=30$ results
in tiny changes around the peaks only, barely visible on the scale of the figure.
Qualitatively, the hopping for $r=3$ resembles the one for half-filling in Fig. \ref{fig:scf2}.
However, the hopping for $r=2$ (which is zero at $q_F=\pi/2$) has a double-peak
structure.

%
%

Returning to the two dominant matrix elements, the question is again, how they compare
with a parabolic law. Near the boundaries, i.e. for small $z$, this can be answered very
simply. The leading contribution to the hopping term is obtained in \eref{tcont2} if the
walker makes only one move, $\ell=s-1$. This gives
\eq{
(\tilde T^{s})_{i,i+1} \simeq z\,s \left(\frac{A}{2}\right)^{s-1} 
\label{tsz1}}
and with the identity
\eq{
\sum_{n=0}^{2m+1} \beta_{m,n} (2m+1-n) = \pi \, 2^{2m}
}
one finds
\eq{
h_{i,i+1} = \pi z \sum_{m=0}^{\infty} \alpha_m A^{2m}= \frac{\pi x(1-x)}{\sin q_F} \, .
\label{hsz}}
Similarly, the diagonal term is obtained if the walker does not move at all, $\ell=s$, which
leads to 
\eq{
h_{i,i} = \frac{-2 \cos q_F \,\pi x(1-x)}{\sin q_F} \, .
\label{dsz}}
Hence, up to a rescaling by the  Fermi velocity, both matrix elements are the same as in $\pi T$.
This corresponds to lifting \eref{epsint3} to the matrix level and means that, close to the 
boundaries, only the small $\eta$ determine the profile.

\begin{figure}[htb]
\center
\includegraphics[width=0.49\columnwidth]{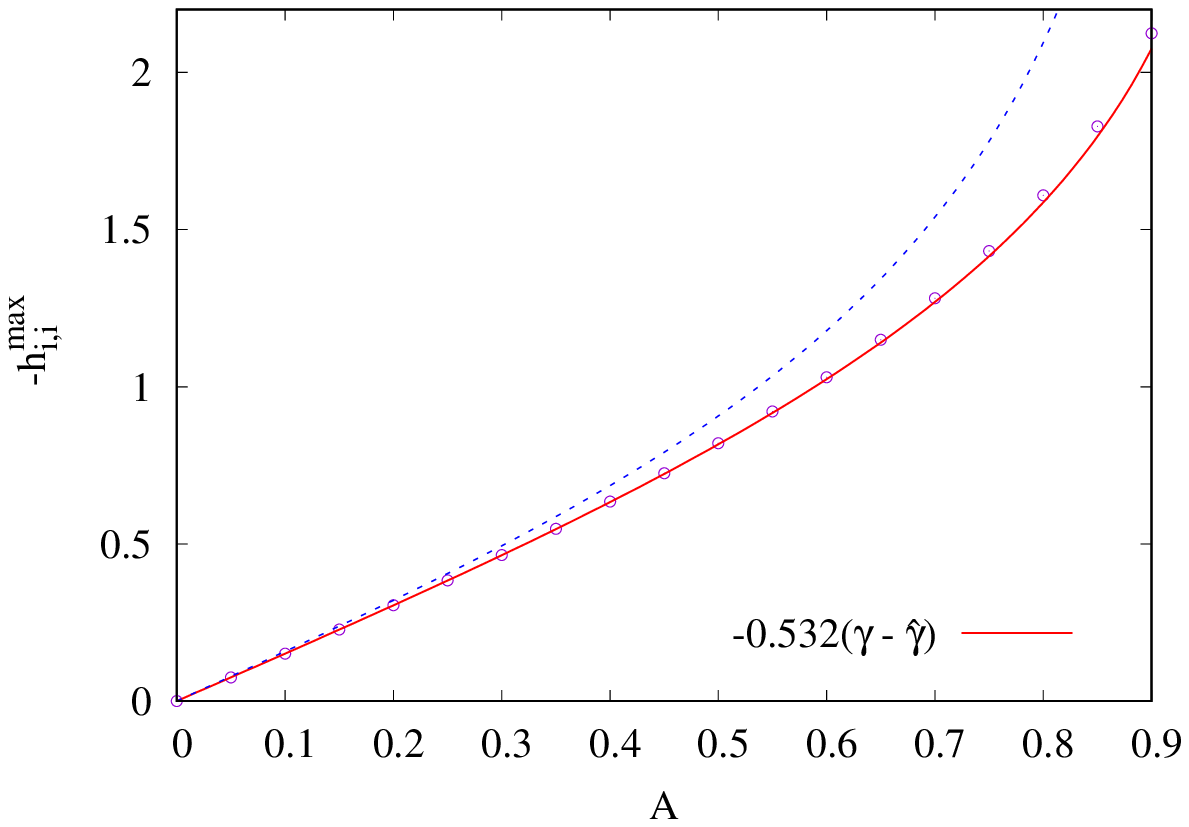}
\includegraphics[width=0.49\columnwidth]{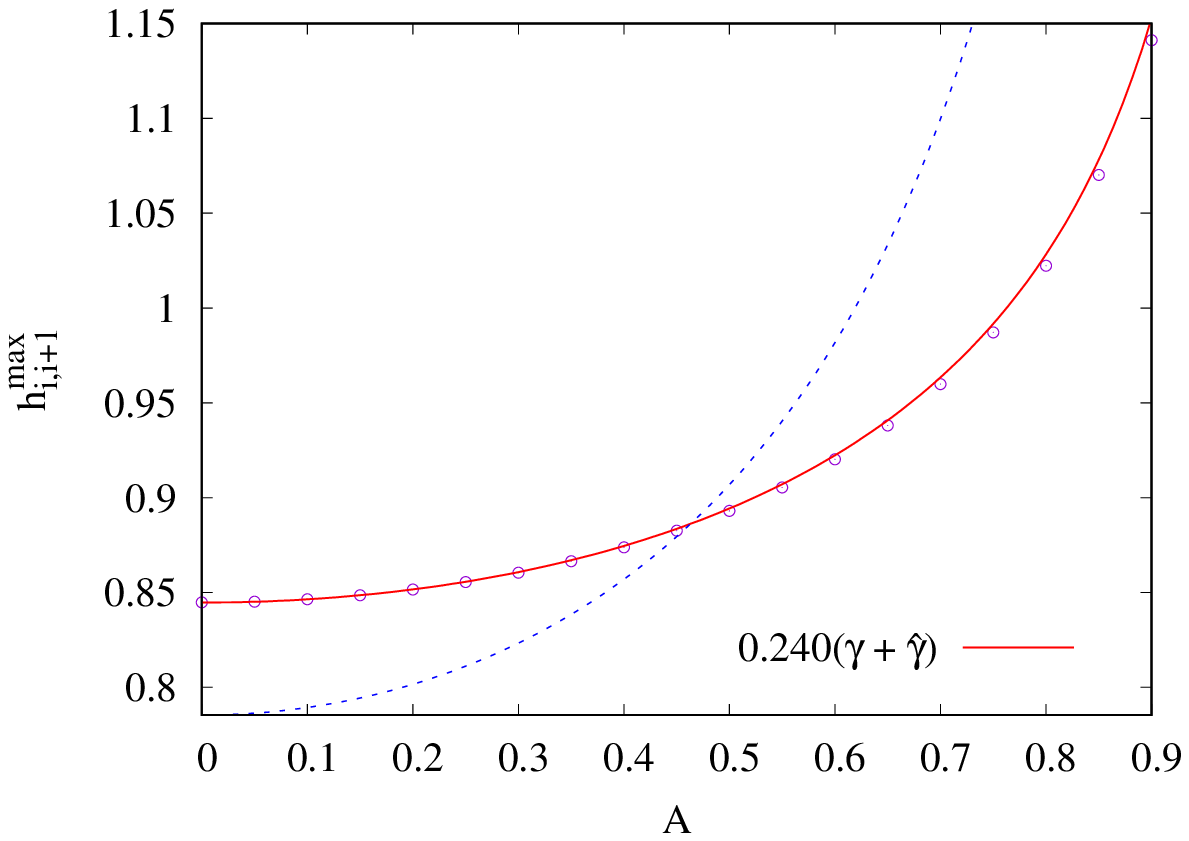}
\caption{Maximal matrix elements of $h$ as functions of the parameter $A=\cos q_F$.
Left: Diagonal elements. Right: Nearest-neighbour hopping.
The full red lines are the approximations \eref{hmax2}, the dotted lines the maxima of the
parabolas \eref{dsz} and \eref{hsz}.}
\label{fig_hmax}
\end{figure}

For the maxima, $z=1/4$, the expression \eref{tcont2} also simplifies. Then the walker is not 
allowed to pause and only the $\ell=0$ term remains. As a result, one obtains for $r=0$
\eq{
h^{\mathrm{max}}_{i,i} = \sum_{n=0}^{\infty} a_{2n+1}A^{2n+1}, \qquad
a_{2n+1} = -\pi \alpha_n \alpha_{2n+1} \,
_3F_2\left(-\frac{1}{4},\frac{1}{4},n+\frac{1}{2};1,n+1;1\right)
\label{hmax0}
}
while for $r=1$ 
\eq{
h^{\mathrm{max}}_{i,i+1} = \sum_{n=0}^{\infty} a_{2n}A^{2n}, \qquad
a_{2n} = \frac{1}{4} \pi \alpha_n \alpha_{2n} \,
        _3F_2\left(\frac{1}{4},\frac{3}{4},n+\frac{1}{2};2,n+1;1\right) .
\label{hmax1}
}
These expressions are shown in Fig. \ref{fig_hmax} for a number of filling parameters $A$.
Whereas $h^{\mathrm{max}}_{i,i}$ is antisymmetric in $A$ and vanishes for half filling, $A=0$, 
the hopping element $h^{\mathrm{max}}_{i,i+1}$ is symmetric and reduces to the expression 
\eref{hscf} at $A=0$, since then only the term $n=0$ survives in \eref{hmax1}. Both amplitudes 
diverge as $A \to 1$. 

A simple approximation for the two quantities is suggested by
looking at the representation \eref{Hij2}. If one assumes that the two pieces scale
roughly as the corresponding extremal eigenvalues $\gamma N$ and $\hat \gamma N$, one can 
write
\eq{
h^{\mathrm{max}}_{i,i}=c_0 \,(\gamma-\hat \gamma)\, , \qquad 
h^{\mathrm{max}}_{i,i+1} =c_1 \,(\gamma+\hat \gamma)\,.
\label{hmax2}
}
With the explicit formula \eref{epsmin2} in Appendix A, one can determine the constants from
the lowest terms in $A$ and then evaluate the right hand sides. This gives $c_0 \simeq 0.532$
and $c_1 \simeq 0.240$ and leads to the solid lines in the Figures. One sees, that they fit the 
asymptotic result very well up to $A \sim 0.9$. 

Comparing now to the maxima of \eref{hsz} and \eref{dsz}, which are shown as dotted lines, one 
sees that the diagonal elements always lie \emph{below} the parabolic law, while the hopping lies
above it for small $A$ (as found for $A=0$ in section 4) and then also below it beyond 
$A \simeq 0.47$ ($q_F \lesssim \pi/3$ or $q_F \gtrsim 2\pi/3$) because $1/\sin q_F$ diverges more strongly.

%
%

\section{Discussion}

We have determined the entanglement Hamiltonian for an interval in a free-fermion
chain. While this can be done with normal numerics for up to $N=20$ sites, larger
systems require special routines and precisely for this case we were able to provide an
analytical treatment. This was done by writing the single-particle eigenvalues $\varepsilon_k$ 
as a power series in the eigenvalues $\lambda_k$ of the matrix $T$, lifting the relation to the
matrix level and calculating the elements of $T^n$ by counting paths.

As a result, we obtained a rather complete picture of $\mathcal{H}$. For half filling, there is
the well-known structure with hopping only between different sublattices, which is largest in the
middle of the system. In a contour plot of the matrix $H$, this longer-range hopping shows
up as a ridge along the anti-diagonal as found in \cite{Arias_etal16} for a slightly non-critical
system. For general filling, there are also diagonal terms and hopping elements within the same
sublattice. The latter show a double-peak structure, which results from negative contributions
in higher powers of $T$. A general filling corresponds to a critical system with Fermi velocity 
$v_F=\sin{q_F}< 1$, and one might have expected that a division of $H$ by $v_F$ is sufficient
to obtain the results from those for half filling. However, while this is true for the low-lying
part of the spectrum and the linear part of the hopping profile, the situation is not so
simple in general.

The deviations from the conformal prediction should be seen as a lattice effect. The longer-range 
hopping is not really contained in a conformally invariant system. In \cite{Arias_etal16}, such
terms were lumped together in order to obtain a local continuum expression and a rather good
agreement with a parabolic law was found. It seems, however, that this is connected with the
slight off-criticality treated there. In our case, all hopping elements for half filling have the
same sign, and combining them would make the deviation from a parabola even larger.

We would like to close with some further remarks.
Firstly, the determination of $\mathcal{H}$ differs in a characteristic way from the usual 
entanglement calculations. For the entanglement entropy, one needs essentially only the low-lying 
$\varepsilon_k$, the higher ones give exponentially small contributions. For $\mathcal{H}$, on the
other hand, the high-lying eigenvalues give the largest contributions,
and in this sense one is sampling different parts of the spectrum in the two cases. 
This was pointed out also in \cite{Arias_etal16}. However, the high eigenvectors are oscillator
functions located in the centre and do not contribute to the linear variation of the nearest-neighbour 
hopping near the boundaries. As noted already in section 5, this behaviour comes from the low-lying 
eigenvalues.

This allows to understand, why the recent approach to interpret the parabolic weight 
factor in $\mathcal{H}$ as a local temperature and to calculate the entanglement entropy from the
thermal equilibrium expression gives the correct asymptotic result \cite{Wong_etal13,Arias_etal16,
Pretko16}. The logarithm $\ln L$ comes from integrating $1/x$, which is the inverse of the linear
slope and thus connected with the low-lying part of the spectrum. Closely related is the so-called 
entanglement contour introduced by Chen and Vidal \cite{Chen/Vidal14,Frerot/Roscilde15,Coser_etal17}. 
In this case, the term $1/x$ can be linked to the envelope $[x(1-x)]^{-1/2}$ of the low-lying 
eigenfunctions $\phi_k$ which enter in the quantity with the largest weight.

Secondly, the commuting matrix $T$ which played a key role in our analysis, is a bit of a mystery, 
because in contrast to $H$ it has exactly the conformal form. Even more astonishingly, the same 
situation is found for a finite ring of $M$ sites. Then such a $T$ also exists and has matrix 
elements which are trigonometric functions \cite{Grünbaum81}
 \begin{eqnarray} 
 \tilde t_i &=&  \sin\left(\pi\frac{i}{M}\right)\sin\left(\pi\frac{N-i}{M}\right), \nonumber \\
 \tilde d_i &=& -2\cos{q_F} \sin\left(\pi\frac{i-1}{M}\right) 
 \sin\left(\pi \frac{N-i}{M}\right) ,
  \label{tridiagonal4} 
 \end{eqnarray}
where, for an odd number of $(2m+1)$ particles, the Fermi momentum is defined as $q_F=(2m+1)\pi/M$.
This is again the conformal form \cite{Wong_etal13,Cardy/Tonni16}. Of course, the diagonal terms are 
only determined up to a global constant, but still the feature is intriguing and one wonders if there
is a deeper reason. 

Thirdly, the case of an interval at the end of a half-infinite system is closely related to the
situation considered here. It can be obtained from a subsystem of $2N+1$ sites in an infinite chain
by restricting oneself to the odd eigenfunctions $\phi_k$ and the corresponding eigenvalues 
\cite{Eisler/Peschel13}. This means that the spacing and the maximal eigenvalues are about twice as 
large as for an interval of length $N$ in the infinite chain. Therefore, with simple numerics, one can 
only reach a size of $N=10$ here. But a commuting matrix $T_s$ exists again and is one-half of the 
matrix $T$ for $2N+1$ sites so that all steps in the asymptotic treatment can be repeated.

\vspace{-0.2cm}

\ack
It is a pleasure to thank Ming-Chiang Chung for help and important remarks in the initial stage of this 
work. We also thank John Cardy for correspondence and Erik Tonni for discussions.
V. E. acknowledges funding from the Austrian Science Fund (FWF) through Lise Meitner Project No. M1854-N36.

\section*{Appendix A: Extremal eigenvalues}

From the results in \cite{Slepian78} one can also find asymptotic expressions for the upper and 
lower ends of the spectra. The $\zeta_k$ closest to one gives$^{\footnotemark}$
\footnotetext{see Eqn. (58) in \cite{Slepian78} for $k=0$ and $2\pi W=q_F$.} 
\begin{equation}
\varepsilon_{min}=-\gamma N +\frac 1 2 \ln N + c
\label{epsmin1} 
\end{equation}
where
\begin{equation}
\gamma= \ln \left(\frac {1+\sin(q_F/2)}{1-\sin(q_F/2)}\right)
       = 2\, \mathrm{artanh\,} (\sin(q_F/2)) \, , \quad  \quad 
  c= \frac 1 2 \ln \left(16\pi \frac {\sin(q_F/2)}{1-\sin^2(q_F/2)}\right) .
\label{epsmin2} 
\end{equation}
The largest eigenvalue follows from the relation \eref{symmetry1} and is 
\begin{equation}
\varepsilon_{max}= \hat \gamma N -\frac 1 2 \ln N - \hat c
\label{epsmax1} 
\end{equation}
where $\hat \gamma$ and $\hat c$ are obtained from $\gamma$ and $c$ via $q_F \to \pi-q_F$ which replaces 
the sines with cosines.
Note that $\gamma$, $\hat \gamma$ can vary between zero and infinity.
For half filling, $q_F=\pi/2$, the result is explicitly
\begin{equation}
\varepsilon_{max}= -\varepsilon_{min}= 1.7627 \, N -\frac 1 2 \ln N - 2.1320 \, .
\label{epsmax2} 
\end{equation}
For $N=40$, for example, both subleading terms are approximately 2 while the leading term is 70.
The logarithmic correction does not appear in the situation treated in Section 4.

Analogous formulae exist for the eigenvalues of $T$. They can be obtained also by taking a 
continuum limit and converting the matrix equation into the differential equation for the
harmonic oscillator. The result is
\begin{eqnarray}
\lambda_{max} &=& \sin^2(q_F/2) - \sin(q_F/2)/N \, , \nonumber \\
\lambda_{min} &=& -\cos^2(q_F/2) + \cos(q_F/2)/N \, .
\label{lambda_ex1}
\end{eqnarray}
Here, there is no logarithmic correction and the constants cannot exceed one. The total width
of the spectrum also equals one to leading order. For half filling, one has
\begin{equation}
\pi N \lambda_{max}= -\pi N \lambda_{min}= 1.5708 \, N - 2.2214 \, .
\label{lambda_ex2} 
\end{equation}
The expressions \eref{epsmax2} and \eref{lambda_ex2} are compared with the 
numerical data in Fig. 1.


\section*{Appendix B: Generalized hypergeometric functions}
The function $_{3}F_{2}$ encountered in the main text is a special case
of the functions $_{p}F_{q}$ which generalize the hypergeometric series to
more parameters in the coefficients. They are defined as \cite{Sneddon61} 
\begin{equation}
_{p}F_{q}(a_1,a_2,...a_p;b_1,b_2,...b_q;z)=
     \sum_{n=0}^{\infty}\; \frac{(a_1)_n(a_2)_n...(a_p)_n}{(b_1)_n(b_2)_n...(b_q)_n}\;
      \frac{z^n}{n!}
\label{gen_hyp} 
\end{equation}
where $(a)_n=a(a+1)(a+2)...(a+n-1)$ is the Pochhammer symbol with the property
\begin{equation}
\Gamma(a+n)=\Gamma(a)\;(a)_n \, .
\label{pochhammer2} 
\end{equation}
In particular, $(a)_1=a$ and $(1)_n=n!$.
The series for $_{p}F_{q}$ converge absolutely for $|z|<1$ and also for $z=1$ if the sum over the
parameters in the denominator is larger than the one in the numerator.
For $p=2,q=1$, they reduce to the normal hypergeometric functions.

To obtain \eref{hpscf}, one proceeds as follows. The explicit expression
for $h_{i,i+2p+1}$ reads
\begin{equation}
h_{i,i+2p+1}= \sum_{m \ge p} 2^{2m}\frac{\Gamma(m+1/2)\Gamma(2m+1/2)\Gamma(2m+2)}
             {\Gamma(m+1)\Gamma(2m+2)\Gamma(m-p+1)\Gamma(m+p+2)}\;z^{2m+1} \, .
\label{element1} 
\end{equation}
Since the summation starts at $m=p$, one shifts the index setting $m=p+n$ with $n \ge 0$.
This gives 
\begin{equation}
h_{i,i+2p+1}= \sum_{n=0}^{\infty} 2^{2n+2p}\frac{\Gamma(n+p+1/2)\Gamma(2n+2p+1/2)}
             {\Gamma(n+p+1)\Gamma(n+1)\Gamma(n+2p+2)}\;z^{2n+2p+1} \, .
\label{element2} 
\end{equation}
The second gamma function in the numerator can be rewritten with the doubling formula
\begin{equation}
\Gamma(2x)=2^{2x-1}\Gamma(x)\Gamma(x+1/2)/\sqrt{\pi} \, .
\label{doubling1} 
\end{equation}
After that, all gamma functions can be expressed by Pochhammer symbols using \eref{pochhammer2}
and one finds
\begin{equation}
h_{i,i+2p+1}= C_p \;z^{2p+1} \sum_{n=0}^{\infty} \frac{(p+1/4)_n (p+1/2)_n(p+3/4)_n}
             {(p+1)_n(2p+2)_n}\;\frac{(4z)^{2n}}{n!}
\label{element3} 
\end{equation}
where the sum is seen to be a $_{3}F_{2}$ function and the prefactor  
\begin{equation}
C_p = 2^{2p} \;\frac{\Gamma(p+1/2)\Gamma(2p+1/2)}{\Gamma(p+1)\Gamma(2p+2)}
\label{prefactor} 
\end{equation}
can be brought into the form given in \eref{hpscf} via the relation 
$\Gamma(p+1/2)=\sqrt{\pi}\;2^{-p}(2p-1)!!$.

Incidentally, also the relation \eref{epslam} can be written in terms of a $_{3}F_{2}$ function.
In this case, the formula is
\begin{equation}
\eta= \pi\lambda \,_{3}F_{2}\left(\frac{1}{4},\frac{1}{2},\frac{3}{4};1,\frac{3}{2};[2\lambda]^2\right) .
\label{epslam_hyp} 
\end{equation}
The results for arbitrary filling involve \emph{sums} over generalized 
hypergeometric functions. For example, \eref{hmax0} is obtained by setting $r=0,z=1/4$ in 
\eref{tcont2} and changing $n \to 2n+1$, since only odd $n$ are allowed, giving
\eq{
h^{\mathrm{max}}_{i,i} = \sum_{m=0}^{\infty} \sum_{n=0}^{m}
\alpha_m \beta_{m,2n+1}
\binom{2m-2n}{m-n} 4^{-(2m-2n)}
\left(\frac{A}{2}\right)^{2n+1} .
\label{hmax3}}
For fixed $n$, the sum starts only at $m=n$. Thus, shifting $m \to m+n$ gives as coefficient of $A^{2n+1}$
\eq{
a_{2n+1} = 2^{-(2n+1)} \sum_{m=0}^{\infty} \alpha_{m+n} \beta_{m+n,2n+1}
\binom{2m}{m} 4^{-2m}
}
and inserting the $\alpha$ and $\beta$, which consist of Gamma functions, one recognizes it as
a $_{3}F_{2}$ function times some factors.
Similarly, the analogue of \eref{epslam_hyp} obtained from \eref{epslam2} is an infinite sum over  
$_{3}F_{2}$ functions.


\pagebreak

\section*{References}

\providecommand{\newblock}{}

\end{document}